\documentclass[final,preprint,5p,times,twocolumn, authoryear]{elsarticle}

\usepackage{amssymb}
\usepackage{subcaption}
\usepackage{amsmath}

\journal{arXiv.org}

\begin{document}
	
	\begin{frontmatter}
		
		

\title{Macroeconomic performance of oil price shocks in Russia}
\author{Kristina Tiron}
\author{Ayaz Zeynalov\cortext[*]{\scriptsize Corresponding author: Ayaz Zeynalov, Prague University of Economics and Business, Faculty of International Relations. Address: , 130 67, Prague, Czechia. T: (+420) 224 095 239; \textit{Email:} ayaz.zeynalov@vse.cz. Kristina Tiron is student at Prague University of Economics and Business, \textit{Email:} tironkris@gmail.com. Financial support from the Prague University of Economics and Business (grant IG212021) is gratefully acknowledged.}}

\affiliation{organization={Prague University of Economics and Business},
	addressline={W. Churchill Sq. 1938/4},
	city={Prague},
	postcode={130 67},
	country={Czechia}}

\begin{abstract}
    Oil price fluctuations severely impact the economies of both oil-exporting and importing countries. High oil prices can benefit oil exporters by increasing foreign currency inflow; however, an economy can suffer from a weakening of the manufacturing sectors and experience a significant downtrend in the country's price competitiveness as the domestic currency appreciates. We investigate the oil price fluctuations from Q1, 2004 to Q3, 2021 and their impact on the Russian macroeconomic indicators, particularly industrial production, exchange rate, inflation and interest rates. We assess whether and how much the Russian macroeconomic variables have been responsive to the oil price fluctuations in recent years. The outcomes from VAR model confirm that the monetary channel is more responsive to oil price shocks than fiscal one. Regarding fiscal channel of the oil price impact, industrial production is strongly pro-cyclical to oil price shocks. As for the monetary channel, higher oil price volatility is pressuring the Russian ruble, inflation and interest rates are substantially counter-cyclical to oil price shocks. 
\end{abstract}




\end{frontmatter}

\section{Motivation}
\label{sec:intro}

The role of oil income in a resource-rich country has been well studied in the literature. Fluctuations of the price in the world oil market directly impact the economies of both exporting and importing countries. In 2022 the world has once more experienced considerable turmoil in the oil market, which impacted the world's economy. Since the 1973 oil price shock, scholars have observed a strong link between the oil price volatility and macroeconomic trends in the world \citep{Hamilton1983}. Factors such as oil demand and supply equilibrium \citep{Hamilton2009, Kim2018, Baek2022}, precautionary demand \citep{Anzuini2015}, speculative activities \citep{Kilian2014}, political uncertainties \citep{Kang2013, Ozdemir2013}, investor sentiment \citep{Qadan2018}, and market-specific factors and financial indicators \citep{Chatziantoniou2021} are the reasons for oil prices volatility. Meanwhile, higher oil price volatility negatively affects stock market returns in developed economies \citep{Diaz2016} and strongly affects the sovereign credit risk of BRICS countries \citep{Bouri2018}. The linkage between oil price shocks and the performance of macroeconomic indicators of oil-producing countries still requires further analysis.

The contribution of this paper is to assess the oil price fluctuations from 2004 to 2021 and their impact on the Russian macroeconomic indicators. The paper's main focus is to study how Russian macroeconomic indicators responded to the oil price shocks after the 2014 sanctions, the 2020-2021 pandemic and the Saudi Arabia oil price war. Russia is the world's second-largest natural gas producer after the United States and the third-largest oil producer after the U.S. and Saudi Arabia and can be referred to as one of the energy superpowers. In 2021, oil and gas accounted for one-third of the total federal budget revenues, representing 60\% of export \citep{MinFin2022}. Considering the production volume and its share in the federal budget, the Russian economy mainly depends on the exports of its natural resources, particularly hydrocarbons. This dependence makes Russia vulnerable to the effects of oil price volatility on the market.

We investigate the relationships between oil prices and Russian macroeconomic indicators utilising the structural vector autoregressive (SVAR) model. The indicators observed are industrial production, exchange rate, inflation rate, and interest rate. The industrial production index (IPI) indicates the influence of the progression of oil prices on Russian industries' output. The analysis is conducted under an assumption that, despite being one of the world's major petroleum producers, Russia cannot influence world oil prices. Our research explains how much the Russian macroeconomic indicators have been responsive to the oil price fluctuations in recent years. We cover the recent economic declines in Russia after 2014 and 2020-2021, which is vital in observing how responsive and dependent the Russian economy is on oil revenues.

\begin{figure}[ht]
	\caption{The progression of the Russian GDP growth and Brent oil prices (2004-2021)}
	\vspace{-5mm}
	\begin{center}
		\includegraphics[width=90mm]{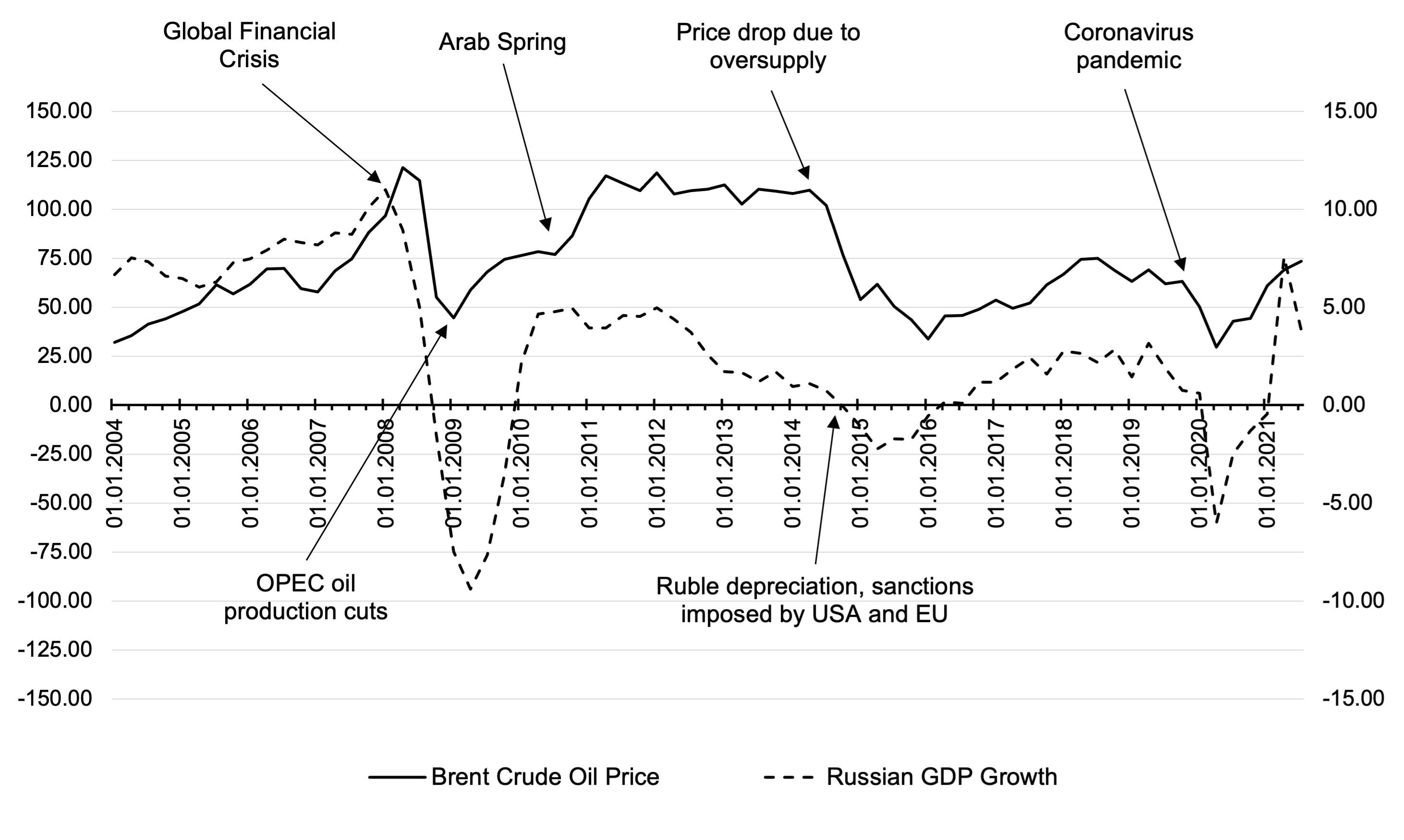}\label{fig:1}
		\begin{tabular*}{0.45\textwidth}{@{\hskip\tabcolsep\extracolsep\fill}ccccc}
			\multicolumn{4}{p{0.45\textwidth}}{\scriptsize \emph{Note:} The authors' estimation is based on data from FRED. The left axis represents Brent oil price; the right axis represents GDP growth.}\\ 
		\end{tabular*}
	\end{center}
\vspace{-5mm}
\end{figure}

High oil prices can benefit the oil exporters as they experience an influx of export revenues, which raises government income and leads to higher government spending and economic growth. On the contrary, economies might suffer a so-called Dutch disease, which indicates a shrinking in non-resource sectors and a significant downtrend in the country's price competitiveness in the market as the domestic currency appreciates. Thus, large oil reserves are sometimes regarded as an economic challenge or a ``curse" rather than a blessing for developing countries. 

Low oil prices reduce oil exporters' revenues, leading to domestic currency depreciation as foreign exchange earnings decline. On the one hand, domestic currency depreciation reduces the affordability of imports. This situation could directly impact domestic companies and households, which are now less capable of purchasing goods and services from abroad, decreasing the population's well-being. On the other hand, the domestic currency depreciation can increase exports, as domestic products now become more competitive in the world markets \citep{Popova2017}. Nevertheless, it is mostly not the case for resource-rich developing countries such as Russia because their producers lack sufficient potential and abilities.

The economic growth of Russia in the period from 2004 until 2021 could be described as highly volatile, yet with clearly defined periods of progress and recession. Figure-\ref{fig:1} exhibits the growth of the Russian GDP from 2004 to 2021. It demonstrates the three economic recession periods of the given time interval. Following a continuous increase in GDP between 2004 and 2008, the first and most significant drop in the Russian economy was in 2008, which corresponded to could global financial crisis. The second economic decline was between 2014-2016, caused by a sharp devaluation of the Russian ruble. The ruble's depreciation was caused by a fall in investors' confidence. The oil market was experiencing a price shock, and sanctions were imposed by the United States and the European Union as a response to the Russian invasion of Ukraine and Crimea annexation. As investors were hurriedly selling off their Russian assets, the ruble's value was distressed, as well as the budget income from oil exports. Finally, the third recession period is the COVID-19 pandemic of 2020, which affected all countries worldwide.

The world has seen several oil price shocks that directly impacted the Russian economy in the past few years. Figure-\ref{fig:1} also represents the Brent Crude oil price and marks the most significant events that caused the trends' oscillations. It is noticeable that the Russian economic recession periods correlate with the oil price distress, and the oil price growth corresponds with the increase in the Russian GDP. It concluded that the two given variables are interconnected, and the world oil prices' trends largely influence the Russian economy. 

Oil pricing is thus a forceful lever that influences the economies of oil-rich countries, especially those that primarily rely on petroleum exports. This calls for an even more meticulous analysis in the case of Russia, taking into account its' oil export dependence and the recent events, including the economic sanctions of 2014 and an oil trade war in 2020. Oil price fluctuations in the market have been primarily reflected in the output of the Russian economy. Periods of an increase in oil prices were also when the Russian economy experienced growth and an influx of revenues. Correspondingly, oil price drops negatively affected the development of the Russian economy reducing its oil export revenues, which represent an essential part of the budget revenues.

The rest of the paper is organized as follows: Section-\ref{sec:bankground} provides an overview of the Russian economy, its position in the world oil market, a description of recent oil price shocks and Russian monetary policy in recent years. Section-\ref{sec:theory} explains the theoretical background, listing the impact channels of oil prices in macroeconomic trends. Section-\ref{sec:method} describes data, methodology and models used. Section-\ref{sec:outcome} presents an empirical outcome using impulse response functions of different macroeconomic indicators to the oil price shocks. In a section-\ref{sec:conclusion}, we summarise results and provide policy implications. Supplementary material and additional results discussed throughout the paper are available in the Appendix.

\section{Russian macroeconomic background}
\label{sec:bankground}

After the collapse of the Soviet Union in 1991, Russia started its shift from a centrally planned economy to a market economy. It adopted responsibility for the external debts of the Soviet Union, which created a great challenge, considering the difference in population and territory between the Russian Federation and the USSR. Overall, the Russian economy performed poorly in the 1990s, resulting in a default on the Russian Central Bank's debt caused by the economic crisis of 1998 and the devaluing of the ruble. According to the International Monetary Fund data, the inflation rate peaked at 1570 \% in 1992, then at 874 \% in 1993. This was caused by a so-called ``shock therapy", a sudden and dramatic change of policies, as the Russian government removed the Soviet price controls. 

The beginning of the 2000s is referred to as the ``golden period" of the Russian economy. The newly elected government implemented several pro-growth reforms to stabilize the economy \citep{Popova2017}. This period is also sometimes called the ``well-fed 2000s", owing to the continuously increasing well-being of citizens. Incomes were growing, as well as consumer expenditures, where increasing demand for imported goods and services. The stability of the domestic currency and continuously rising oil prices, which reached their historical maximum during the summer of 2008, have played an essential role in the significant growth of the Russian economy during that period. 

The 2008 financial crisis caused a steep decline in oil prices as the demand for petroleum by consumers and businesses decreased significantly. The Global Financial Crisis is a period of severe stress in global financial markets and banking systems between the middle of 2007 and the beginning of 2009. A crash in the US housing market caused global economic disruptions, and most world economies experienced deep recessions. Consequently, the energy demand drastically fell, lowering oil and gas prices. Oil prices fell from around 130 US\$ to less than 40 US\$ in just half a year, which directly affected the oil-exporting countries and companies due to a significant downtrend in energy revenues. During the 2008 crisis, Russian economic performance, including the mining sector, negatively responded to a shock in oil prices \citep{Balashova2020}. The GDP growth during that period attained negative values, signifying a recession period for more than a year. However, \citet{Yoshino2016} shows that the Russian economy recovered faster after the crisis owing to highly effective measures applied by the government.

In order to stabilize the oil market and its prices after the 2008 Global Financial Crisis, OPEC countries pledged to cut oil production for the first time in eight years. OPEC members' oil supply reductions were settled to be 4.2 million barrels per day, with its largest producer Saudi Arabia implementing the most significant production limitations. Consequently, oil prices rose, reaching a price of more than 70 US\$ per barrel. A positive oil price shock occurred in the first half of 2011, reaching from 80 US\$ to 120 US\$ per barrel, sufficiently increasing Russia's petroleum export revenues. The oil price rise appeared for several reasons, including the Arab Spring, Libya's civil war, Japan's nuclear power plant explosion, etc. \citep{Popova2017}. The oil price volatility is attributed to the region's widespread fears concerning supply disruptions. Consequently, this led to a rise in oil and gas revenues in Russia, with its value rising from 3,8 trillion rubles to 5,6 trillion \citep{MinFin2022}. The Russian federal budget revenue details (Figure-\ref{fig:a1}), and crude oil and oil products export details (Figure-\ref{fig:a2}) are presented in Appendix.

\begin{figure}[ht]
	\caption{The progression of the Russian Monetary policy indicators and Brent oil prices (2004-2021)}
	\vspace{-5mm}
	\begin{center}
		\includegraphics[width=90mm]{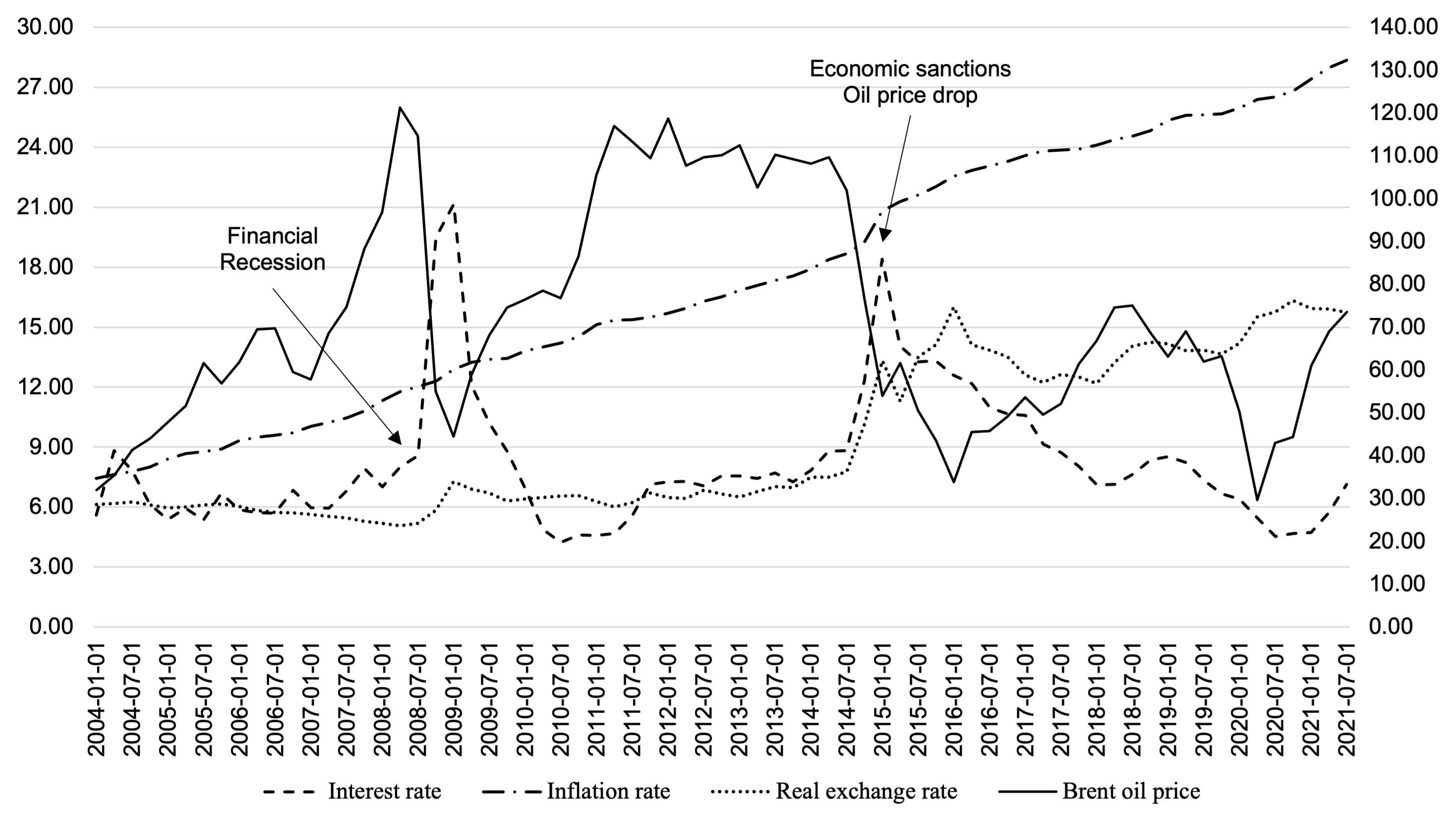}\label{fig:2}
		\begin{tabular*}{0.45\textwidth}{@{\hskip\tabcolsep\extracolsep\fill}ccccc}
			\multicolumn{4}{p{0.45\textwidth}}{\scriptsize \emph{Note:} The authors' estimation is based on data from FRED. The left axis represents interest rate; the right axis represents inflation and exchange rates, and Brent oil price.}\\ 
		\end{tabular*}
	\end{center}
\vspace{-5mm}
\end{figure}

The volatility of the world oil prices directly impacts Russian monetary policy indicators. Russian interest rate is counter-cyclical; once oil price increases, the interest rate is reduced, and vice versa (Figure-\ref{fig:2}). It emphasizes that higher oil price leads to higher foreign reserve inflows to the domestic market, therefore pressuring the interest rate downward. On the contrary, lower oil prices lead to lower income, increased demand for borrowing, and an increased interest rate. Interest rate reached its highest during the financial crisis of 2008; the second highest occurred in 2014 during economic sanctions and dramatic oil price reduction. The real exchange rate was stable till 2014 despite high oil revenue inflow; however, it depreciated after 2014, triggered by sanctions and oil price drops. Inflation steadily increased 32 \% from 2015 to 2021 (Figure-\ref{fig:2}).

Following that, in 2014, the world witnessed a drastic drop in the oil prices as the US and Canadian shale oil (tight oil) production vastly expanded its market share, along with increasing oil supply from production growth in Saudi Arabia, Libya, and Iraq \citep{Nyangarika2018}. Also, a weakening of the Chinese economy during that period caused a fall in demand across commodities markets. The drop was rapid, and oil prices decreased from more than \$100 per barrel to almost 50 US\$ per barrel. The fall continued in 2015, reaching fewer than 40 US\$ per barrel. By and large, the oil market from 2014 to 2016 is characterized by extreme volatility. This oil price drop caused a recession in GDP growth in Russia. This shock also affected the Russian exchange rate, causing a depreciation of the Russian ruble \citep{Alekhina2019}. The oil price shock's impact on the Russian domestic exchange rate is explained by the fact that petrol is primarily sold as foreign currency. The world oil price is also believed to be the most significant external factor influencing the Russian economy's US dollars to ruble ratio \citep{Nyangarika2018}. A worldwide reaction to the Russian geopolitical decisions also is a cause of a weakening of the Russian currency and economic downtrend – the US and the EU imposed sanctions on Russia as a response to the invasion of Ukraine and the annexation of Crimea.

The most recent oil price shock was initially provoked by a severe decrease in demand as the global market slowed down due to the coronavirus pandemic in 2020. As most countries have imposed travel and export restrictions, banned commercial activities, etc., the world has fallen into an economic recession. Industrial production across the globe has drastically declined, lowering the energy demand and thus decreasing world oil prices. The drastic fall in world goods and services trade as the COVID-19-related restrictions constrained economic activities worldwide. With the declines in international trade and overall production worldwide, the demand for crude oil decreased, lowering its market price.

Subsequently, in early March 2020, Saudi Arabia began a price war with Russia by notifying its buyers that OPEC was planning on increasing oil output. It was a response to Russia's rejection to cut oil production to stabilize crude oil markets and raise oil prices, as the US shale oil production was expanding. Moreover, oil demand was falling globally in light of the global economic crisis caused by the pandemic. The production cut resulted in the oil price plummeting, reaching its lowest since 2002 and falling below 20 US\$ per barrel. The fall of the oil prices on the global market was the largest since the Gulf War, erupting at the beginning of the 1990s \citep{Gupta2020}. The oversupply of oil in the market even made the prices of West Texas Intermediate (WTI) futures reach negative values for the first time in history. As the marginal production costs of oil in the Russian Federation are significantly higher than those of the OPEC countries, it experienced a considerable negative impact from the price war. The world oil price fell below the cost of producing and transporting the oil, making Russian oil companies lose substantial revenues. Later, in April 2020, Russia and Saudi Arabia reached a cut agreement. However, the negative impacts of the outbreak were more durable than the positive impacts generated during the truce \citep{Ma2021}. Nevertheless, oil prices increased by the end of 2020 due to production cuts and the worldwide vaccination roll-out. They were also easing the restrictions preventing COVID-19 from spreading, raising optimism and stimulating global markets. 

Due to uncertainties during the conflict, the Ukrainian-Russian war triggered the oil price increase. In the post-pandemic period, the global economy had already been facing different economic challenges: higher inflation, negative supply shock as production throughout the world drastically plummeted, negative demand shock as consumer behaviour changed, and slowed down manufacturing activity due to global supply chain disruptions. Additionally, Russia has been facing severe sanctions in response to its aggression. However, higher oil prices can ease the negative impacts of the sanctions.

\section{Theoretical background}
\label{sec:theory}

As for most other commodities, the global market determines the price of oil. The overall state of the global economy significantly impacts fluctuations in oil prices. When economies are booming, production is at total capacity, increasing the demand for energy, which increases oil prices. On the contrary, industrial production decreases during economic recessions and crises, and countries demand less petroleum, lowering their market price. Nevertheless, such cartels as OPEC, being the biggest and the most influential organization in the sector, can also influence the oil prices by cutting its production in order to, for example, maintain a certain level of oil price.

\begin{figure}[ht]
	\caption{The channels of macroeconomic response to the oil price increase}
	\vspace{-5mm}
	\begin{center}
		\includegraphics[width=80mm]{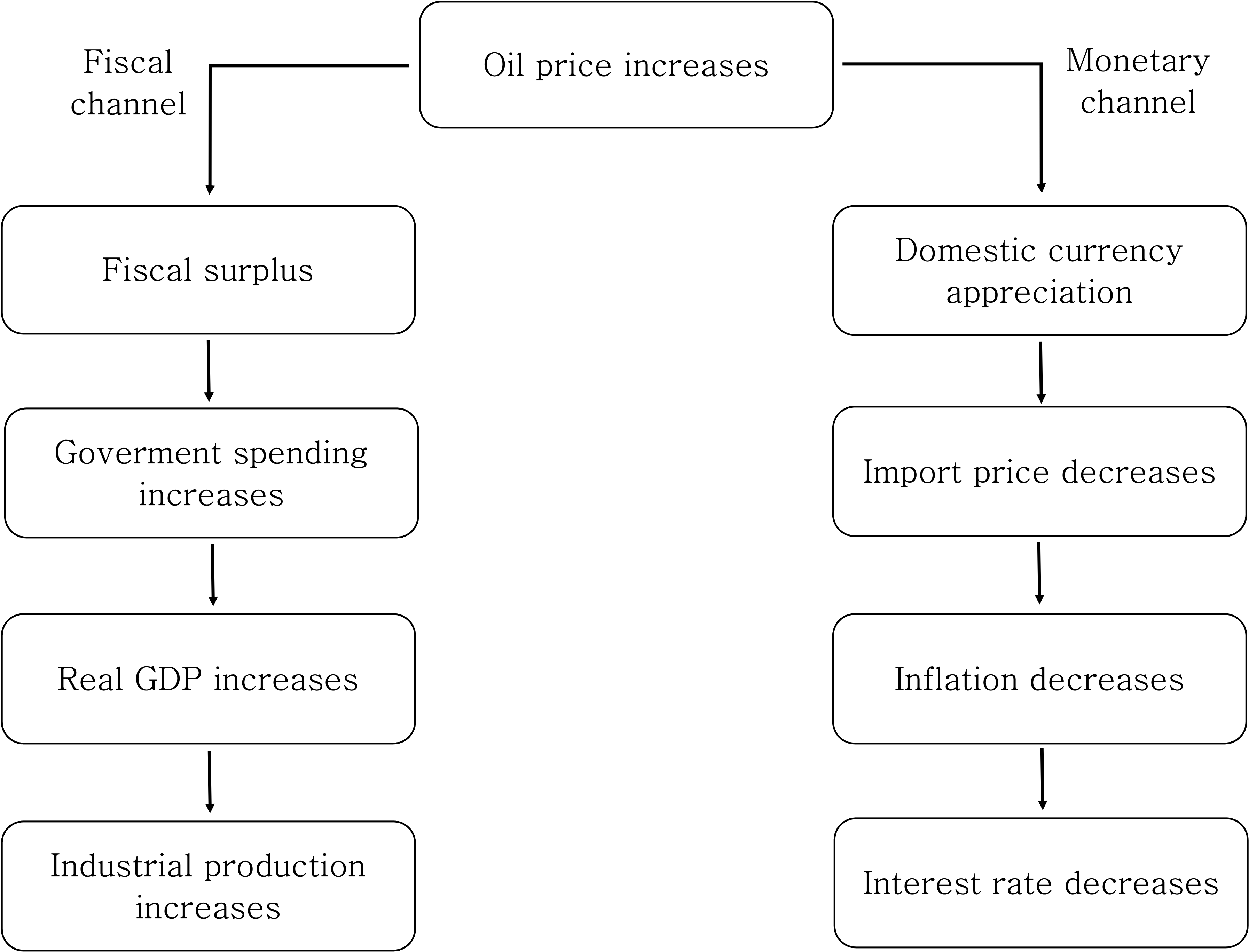}\label{fig:3}
	\end{center}
	\vspace{-5mm}
\end{figure}

Supply and demand equilibrium is not the only factor that impacts crude oil price fluctuations. Oil markets are also driven by oil futures contracts, binding agreements granting their owners the right to buy petroleum at a prearranged price. Different circumstances can lead to an unexpected change in oil pricing or an oil price shock. \cite{Kilian2009} describes three types of oil price shocks: supply shock, aggregate demand shock, and precautionary demand shock. The supply shock indicates a shock to the current availability of the oil. If there occurs a sudden disruption of oil production, it causes the oil price to increase and also the production in other regions to grow in order to meet the amounts of demand. Aggregate demand shock is caused by the volatility of the global business cycles. An example of such a shock could be the sharp decline in oil demand during the 2020 coronavirus pandemic, as the global markets were shutting down and productions throughout the globe decreased. The third shock is precautionary demand shock, caused by expectations and uncertainties in the oil market. An example of such a shock could be the increase in the oil price during the Arab Spring in 2011. The protests, which erupted in several countries in the Middle East and North America, the main oil production regions, raised concerns about possible oil supply disruptions. 

Since crude oil is one of Russia's most significant exported commodities, fluctuations in the oil market can influence the country's economy. Oil prices affect the Russian economy through fiscal and monetary channels, as depicted in Figure-\ref{fig:3}. With an augmentation of the oil price on the market, a larger capital in the form of foreign currency inflows into the economy, causing a domestic exchange rate appreciation, which for its part, decreases prices for imported goods and services. Since imports create a significant portion of total consumer goods, with an increase in oil prices, the Russian economy experiences deflation, with the general level of prices decreasing. Then, according to the Taylor rule's recommendation, interest rates are reduced. From a fiscal channel point of view, the Russian economy benefits from the oil price increase in the form of taxes on the export of energy resources. This creates a fiscal surplus, raising government spending and increasing real GDP.

In accord with \cite{Ito2012} study, oil price shocks and Russia's economic growth have a positive interrelationship, meaning that an increase or decrease in oil prices contributes to the growth or decline of the country's GDP. In particular, the effect of oil price shock on the GDP growth rate of an oil-exporting country is expected to be positive due to an increase in revenues \citep{Taghizadeh2019}. This effect is more significant in bigger economies in terms of oil export, such as the Russian Federation, UAE, Saudi Arabia, etc. According to the \cite{Yoshino2016} study, high oil prices positively and significantly affect Russia's inflation rate. \citet{Oloko2021} claim that monetary policies will adapt to the oil price shock if a country has either floating regimes and inflation targeting or pegged regime and non-inflation targeting monetary policy.

Nevertheless, oil-rich countries might as well experience the negative impacts of the price increase, also called Dutch disease. It indicates a negative relationship between the rise in the price of natural resources and the economic growth of a particular country. According to \cite{Otaha2012}, the influx of revenues from oil exports collapses other sectors of the economy. Such a situation can be observed in oil-rich developing countries like Iran or Nigeria, which remain highly dependent on petroleum exports. Since the beginning of oil extraction in this country, its economy has drastically declined \citep{Otaha2012}. The rapid growth of the oil sector also entailed a collapse of other sectors of the Nigerian economy, such as agriculture and manufacturing. 

Russia is also suffering from such a disease. As the petroleum industry dominates the economy and is an essential source of revenue for the federal budget, other sectors have lacked incentives for development \citep{Popova2017}. A large portion of the foreign direct investments has been targeting the oil and gas sector, preventing other sectors from expanding. Russian macroeconomic variables are dependent on an exogenous unstable factor. This has forced the Russian government to implement measures and policies that effectively stabilize the economy after an oil price shock. In 2004, after a sudden increase in oil prices, the Russian Federation established the Stabilization Fund, which is considered Russia's most prominent policy against oil crises \citep{Popova2017}. This fund aimed to accumulate and reinvest oil revenues into foreign bonds to provide and support the federal government budget. Later, in 2008 this fund was divided into two separate funds: the Russian National Wealth Fund and the Reserve Fund.

Considering a strong relationship between the world oil prices and the Russian macroeconomic performance, some studies, such as \cite{Ponka2019} and \cite{Balashova2020}, suggest that oil price fluctuations can help predict changes in the growth rate of Russia's main economic activity. Oil prices are procyclical and lead the economic activities in Russia and thus making it a valuable indicator of possible future recession periods. Scholars have made different suggestions and recommendations for Russian officials to mitigate the harmful dependence of the Russian economy on the so-called ``black gold". The \cite{Alekhina2019} study emphasizes how an energy-exporting country must diversify the range of its revenues from too much reliance on oil export. Also, a reduction of dependence on energy resources is advised to achieve through transitioning from an industrial economy to an innovative one by improving the investment climate in the country for foreign investors.  

\section{Data and Methodology}
\label{sec:method}

\begin{table}[ht]
	\caption{Descriptive statistics\label{tab:1}}
	\vspace{-5mm}
	{\footnotesize
		\begin{center}
			\begin{tabular}{p{6mm} p{32mm} p{35mm} }		
				\hline
				$ipi_t$     &	Industrial production index	&	Production of Total Industry, 2015=100	\\
				$\pi_t$     &	Inflation rate	&	CPI inflation rate, 2015 base year	\\
				$i_t$       &	Interest rate	&	3-Month or 90-day Rates and Yields	\\
				$e^{\frac{rub}{usd}}_t$  &	Exchange rate	&	Ruble-US dollar exchange rate	\\
				$p^{oil}_t$ &	Oil price	&	Price of Brent crude oil, dollar per barrel	\\
				$\Delta y$  &	GDP growth	&	Gross Domestic Product in Constant Prices	\\
				\hline
			\end{tabular}
			\begin{tabular*}{0.45\textwidth}{@{\hskip\tabcolsep\extracolsep\fill}ccccc}
				\multicolumn{4}{p{0.42\textwidth}}{\scriptsize \emph{Source:} FRED Archival Economic Data, Ministry of Energy of Russia, Russian Federal State Statistics Service. Data for IPI are seasonally adjusted.}\\ 
			\end{tabular*}
	\end{center}}
	\vspace{-5mm}
\end{table}

To analyze the interdependence of the oil price volatility and the economy of Russia, we used quarterly data from Q1 2004 to Q3 2021, which creates 71 observations for each of the chosen variables. This period includes three main oil price shocks of the last two decades: the 2008 Financial Recession, the oil market's oversupply in 2014, and the COVID-19 pandemic and oil price war between Saudi Arabia and Russia in 2020. The macroeconomic variables chosen for this analysis can be observed in Table-\ref{tab:1}.

In order to assess the response of the Russian macroeconomic indicators to the world oil price fluctuations, we have chosen four macroeconomic indicators: the Industrial Production Index (IPI), the exchange rate, the Consumer Price Index (CPI) and the interest rate. The fiscal channel is explained by the Industrial Production Index (IPI), which is measured as an indicator of economic output performance. It measures the real output in the industrial production (mining, manufacturing and energy sectors) relative to a base year. This variable represents changes in business cycles in the country. As the industrial sector still plays a vital role in the Russian economy, we found it highly important to use the production index in our analysis. Besides total manufacturing output, it also measures production capacity levels, an estimated production volume that could be maintained in a country. Being an index, this variable represents a change expressed in percent relative to the base year. The base year for this study is 2015.

The monetary channel is explained by currency appreciation. We used the Ruble to US Dollar exchange rate since oil-exporting countries settle deals mainly in  US dollars. As crude oil is traded in US dollars, its price fluctuations on the market directly impact the domestic currency of Russia. The inflation rate evaluates the rate at which the level of prices rises in the economy, marking a decline in purchasing power of a currency. The chosen inflation index for this analysis is the Consumer Price Index (CPI), which measures an annual percentage change in the average price of a consumer basket consisting of goods and services. It will serve as an indicator of an increase in the cost of living in the economy. 

The interest rate is an amount a lender charges on the amount loaned to the borrower. It is also a vital monetary policy tool used by the Central Banks of countries to mitigate the inflation rates and influence the nation's economic activity by raising or reducing investment and consumption in the economy. We use the 3-Month or 90-day Rates and Yields, or Interbank rate, which represents an interest charged on short-term loans between financial establishments.

As an indicator for the oil price, we used Brent Crude oil prices, as it is the most widely used global oil price benchmark and because the Russian export oil prices are pegged to this oil blend. GDP growth is measured in constant prices.

We used the Vector Autoregressive (VAR) method to study the impact of oil price volatility on the Russian macroeconomic indicator. The method was developed by \citet{Sims1980} and is based on the Granger Causality test \citep{Granger1969}. The difference between the VAR model and other auto-regressive models is the fact that it predicts bi-directional influences between the time series. The standard Vector Autoregressive model is expressed below:  
\begin{equation*}
    y_t=\alpha_1 + \sum_{i=1}^{p} \delta_{i} y_{t-i} + \sum_{j=1}^{p} \nu_{i} x_{t-j} + u_{t}
\end{equation*}
\begin{equation*}
    x_t=\beta + \sum_{i=1}^{p} \gamma_{i} x_{t-i} + \sum_{j=1}^{p} \mu_{i} y_{t-j} + \epsilon_{t}
\end{equation*}

VAR model has advantages due to the fact, that macroeconomic indicators are linear functions of the past values of themselves and each series is a linear function of the past values of other macroeconomic variables. For our analysis of the influence of the oil prices to the macroeconomic indicators, we estimated four linear regression models with different dependent variables. The four models correspond with the four macroeconomic indicators:

\begin{eqnarray*}
    lnipi_t= \alpha_0 + \alpha_1 lnipi_{t-1} + \alpha_2 lncpi_{t-1} + \alpha_3 lnreer_{t-1} + \\ +  \alpha_4 ir  + \alpha_5 lnoilp_{t-1} + \epsilon_t
\end{eqnarray*}
\begin{eqnarray*}
    lnreer_t= \beta_0 + \beta_1 lnreer_{t-1} + \beta_2 lnipi_{t-1} + \beta_3 lncpi_{t-1} +  \\ + \beta_4 ir + \beta_5 lnoilp_{t-1} + e_t
\end{eqnarray*}
\begin{eqnarray*}
    lncpi_t= \delta_0 + \delta_{1} lncpi_{t-1} + \delta_2 lnipi_{t-1} + \delta_3 lnreer_{t-1} +\\ +  \delta_4 ir +  \delta_5 lnoilp_{t-1} + u_t
\end{eqnarray*}
\begin{eqnarray*}
    ir_t= \gamma_0 + \gamma_1 ir_{t-1} + \gamma_2 lnipi_{t-1} + \gamma_3 lnreer_{t-1} +  \gamma_4 lncpi_{t-1}  \\ + \gamma_5 lnoilp_{t-1} + \eta_t
\end{eqnarray*}

We assume that the oil prices are exogenously given to the Russian Federation, and thus Russian macroeconomic indicators do not possess the possibility of influencing the world oil prices. The next assumption is also that the variables in the model influence each other and also are influenced by their past values. The term $ln$ is the natural logarithm of the variables. The logarithm form of the variables is chosen in order to convert the values of the indicators into percentage change. The interest rate is the only variable not converted to a logarithm form, as it is already expressed in levels. The parameters $ipi$, $reer$, $cpi$, $ir$, and $oilp$ indicate the following variables: industrial production index, real exchange rate, consumer price index (inflation rate), interest rate and Brent oil price. The terms $\alpha_i$, $\beta_i$, $\delta_i$ and $\gamma_i$ represent the response coefficients. The terms $\epsilon_t$, $e_t$, $u_t$ and $\eta_t$ are random errors, that represent the uncertainty in models. If the actual value of the dependent variable does not correlate with the value that the model predicts, the error term does not equal zero and indicates the existence of other factors that might be influencing the dependent variable.

\section{Results}
\label{sec:outcome}

\subsection{Unit root}

Analyzing time series can sometimes be limited by unpredictable patterns, making it harder to assess the data statistically. In order to prove that the changes of a variable over time are not randomized, it is necessary to check the time series for stationarity. Stationarity refers to the statistical properties that generate a time series, such as mean and standard deviation, which do not change over time. In order to eliminate the possibility of the series used in the analysis being non-stationary, we used the unit root tests that determine an existence of a unit root in a time series. A unit root is a stochastic or random trend of a variable, a random and unpredictable pattern, meaning a time series with a unit root is non-stationary.

To assess the stationarity of the given variables and the world oil price, we used an Augmented Dickey-Fuller (ADF) test and Kwiatkowski-Phillips-Schmidt-Shin (KPSS) unit root tests. The variables (except the interest rate) were converted to their logarithmic form. Since the ADF test has a low power to reject the null hypothesis, the KPSS test is used to confirm the results. The null hypothesis for the ADF test is that a unit root exists in the time series, meaning it is non-stationary. Thus it is not recommended to use the variable for the empirical analysis. The KPSS test's null hypothesis is the opposite, meaning there is no unit root. We ran these tests using the variables in levels and then their first differences, a change from one period to the next.

\begin{table}[htp]
	\caption{Unit root tests, Q1 2004 - Q3 2021\label{tab:2}}
	\vspace{-5mm}
	{\footnotesize
	\begin{center}
	\begin{tabular}{p{15mm} l l | p{15mm} c c }	
	\hline
    Level	        &	ADF 	&	KPSS 	&	First 	&	ADF 	&	KPSS 	    \\
        	           &	 test	&	 test	&	 difference	&	 test	&	 test	    \\
    \hline
    $log(p^{oil}_t)$&	-2.668	    &	0.844	    &   $\Delta log(p^{oil}_t)$     &	 -6.994*	    &	0.074*	\\
    $log(ipi_t)$	&	-1.122	    &	0.286	    &   $\Delta log(ipi_t)$	        &	 -6.739*	    &	0.059* 	\\
    $Growth$	&	-2.272 	    &	0.275	    &   $\Delta Growth$       &	 -6.093*	    &	0.043* 	\\
    $log e^{\frac{USD}{RUB}}$	    &	-0.194	    &   0.863   &	$\Delta log e^{\frac{USD}{RUB}}$&	 -7.582*	&	0.083*	\\
    $log(\pi_t)$	&	-3.732*	    &	1.350	    &	$\Delta log(\pi_t)$	        &	 -5.453*      &	0.065*	        \\
    $i_t$	        &	-2.869	    &	0.351	    &	$\Delta i_t$                &    -7.356*	    &	0.031*	        \\
	\hline
	\end{tabular}
	\begin{tabular*}{0.48\textwidth}{@{\hskip\tabcolsep\extracolsep\fill}ccccc}
	\multicolumn{4}{p{0.45\textwidth}}{\scriptsize \emph{Note:} *The selected time series is stationary at the 1 \% significance level}\\ 
	\end{tabular*}
	\end{center}}
	\vspace{-5mm}
\end{table}

\begin{table*}[htbp]
	\caption{VAR Granger causality test (Russia, Q1 2004 - Q3 2021)\label{tab:3}}
	{\footnotesize
		\begin{center}
			\begin{tabular}{p{45mm} | p{14mm} p{15mm} | p{14mm} p{15mm} | p{14mm} p{15mm} |}	
				\hline
				& \multicolumn{2}{c}{2004-2021}	& \multicolumn{2}{c}{2004-2008}     & \multicolumn{2}{c}{2008-2021}     \\
				\hline	                                
				Null hypothesis                     & $\chi^2$      & p-value       &   $\chi^2$        &  p-value     & $\chi^2$      & p-value         \\
				\hline
				$P^{oil}_t$ does not cause $ipi_t$  &  9.911        & $0.007^{**}$  &   1.171           &   0.383       &  3.235        & $0.049^{**}$      \\
				$P^{oil}_t$ does not cause $reer_t$ &  4.344        &  0.114        &   0.382           &   0.700       &  1.238        &  0.300            \\
				$P^{oil}_t$ does not cause $\pi_t$  &  1.871        &  0.392        &   2.936           &   0.144       &  0.120        &  0.887            \\
				$P^{oil}_t$ does not cause $i_t$    &  2.960        &  0.228        &   4.896           &  $0.066^{*}$  &  1.530        &  0.229            \\
				\hline
				$ipi_t$ does not cause $P^{oil}_t$  &  1.232        &  0.540        &   0.094           &   0.911       &  0.509        &  0.605            \\
				$reer_t$ does not cause $P^{oil}_t$ &  0.090        &  0.956        &   0.059           &   0.942       &  0.020        &  0.981            \\
				$\pi$ does not cause $P^{oil}_t$    &  10.364       & $0.006^{**}$  &   1.004           &   0.431       &  3.108        & $0.055^{*}$       \\
				$i_t$ does not cause $P^{oil}_t$    &  1.232        &  0.231        &   0.142           &   0.871       &  1.479        &  0.240            \\
				\hline
			\end{tabular}
			\begin{tabular*}{0.90\textwidth}{@{\hskip\tabcolsep\extracolsep\fill}ccccc}
				\multicolumn{4}{p{0.87\textwidth}}{\scriptsize \emph{Note:} Variables are in first log-difference forms, except $i_t$ is in first-level difference.  $**$ and $*$ represent the rejection of the null hypothesis at the 5\% and 10\% level of significance, respectively.}\\ 
			\end{tabular*}
	\end{center}}
	\vspace{-5mm}
\end{table*}

The results of the tests are presented in Table-\ref{tab:2}. As it can be observed in the table, the starred results are stationary at a 1\% significance level. It is notable that besides the inflation variable, all the data series are non-stationary and thus unsuitable for the analysis. It is important to note that the ADF test has a relatively high Type I error rate, which means that rejecting a null hypothesis might be incorrect. This can be the case in the results of the inflation rate, which according to the ADF test, is stationary. We have checked the KPSS test results to verify this outcome, which indicates that the inflation rate time series is non-stationary only in its first difference. As a result of ADF and the KPSS tests, we use the first differences of logarithms of the variables, as they were proven to be stationary time series and thus cannot be influenced by randomness.

\subsection{Granger causality}

In order to assess the causality direction between Brent crude oil prices and the macroeconomic indicators, notably the Industrial Production Index and Consumer Price Index, we used the Granger causality test for the full sample and two sub-samples: 2004-2008; 2008-2021. The Granger causality allows examining whether a given time series is capable of forecasting another, which is questioned in the null hypothesis. If one variable Y is proved to “Granger-cause” the other variable X, it means that the past values of the Y contain information that might be useful in estimating the values of X. The statistical significance of the analysis results is based on a p-value. One can reject the null hypothesis if the probability value is less than a 5\% significance level.

The selected results of the Granger test are represented in Table-\ref{tab:3}. The first null hypothesis suggests that the world prices of oil do not cause a change in industrial production in Russia. The results are statistically significant for the full sample and the period after 2008; thus, we reject the null hypothesis. It confirms that by observing changes in the world oil prices, one is able to predict the changes in the Russian Industrial Production Index or in the output of the Russian industry sectors. From the first quarter of 2004 to the third quarter of 2008, the oil prices did not Granger-cause Russia's production index, whereas in the period after 2008, the production index was affected by changes in the oil price.

The second null hypothesis implies that the oil price volatility does not cause changes in the exchange rate of the Russian economy. We can not reject the null hypothesis for the full sample or for two different sub-samples. The results are very similar to the third hypothesis regarding the role of oil prices on the inflation rate. In the hypothesis of whether oil prices cause changes in interest rate, we reject causality for the full sample, but it is significant only for the period from 2004 to the third quarter of 2008. It shows that one of the determinants of interest rate level was the oil price during the pre-2008 financial crisis period. However, the causality is not strong, and the rejection of the null hypothesis is possible only at a 10\% level of significance.

As an outcome of the test results of the causality direction from the Russian macroeconomic indicator to the world oil prices, we have found that only inflation significantly impacts both the full sample and the post-2008 Financial crisis period. During the period from 2008 to 2021, the biggest shock to the inflation rate was the Russian economic crisis of 2014, as the US and the EU imposed economic sanctions against the Russian Federation. The economic sanctions and fall in the oil prices during that period caused a severe capital outflow and a depreciation of the Russian domestic currency. It all contributed to the increase in the Consumer Price Index in the following year.

\subsection{Impulse response analysis}
In order to describe and depict the evolution of the variables in the VAR model and to further assess the tendencies of Granger causality results, we used the Impulse Response Function (IRF) analysis. The IRF analysis describes the progression of the variables in reaction to a shock or a change in one or more of the variables in the model along a specified time horizon. It allows for assessing the transmission of a shock within a system of equations and is an essential tool in empirical causal analysis. A shock to a dynamic system is an impulse, or a brief input signal, that causes an impulse response. In other words, a change in a variable is, to a certain degree, passed to other variables. IRF traces the impacts on present and future values of the dependent variable caused by a one standard deviation shock of another variable in a system. We estimated responses of the variables to a one standard deviation shock for eight consequent periods, quarters.

\begin{figure}[htbp]
	\caption{Response of Russian macroeconomic indicators to oil shocks \label{fig:4}}
		\vspace{-5mm}
	\begin{center}
	\includegraphics[width=65mm]{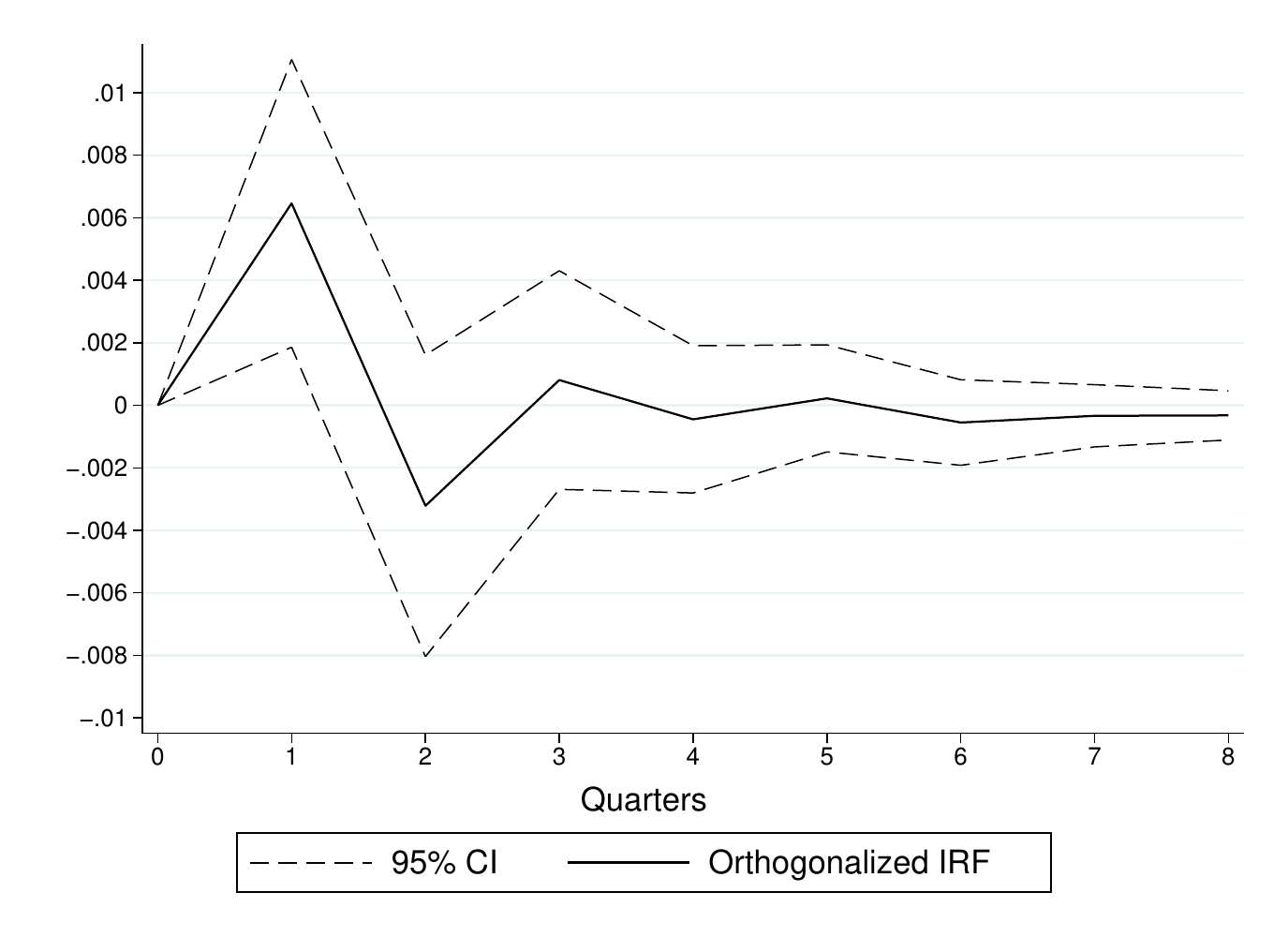}
	\subcaption{Response of industrial production to oil shocks}
	\includegraphics[width=75mm]{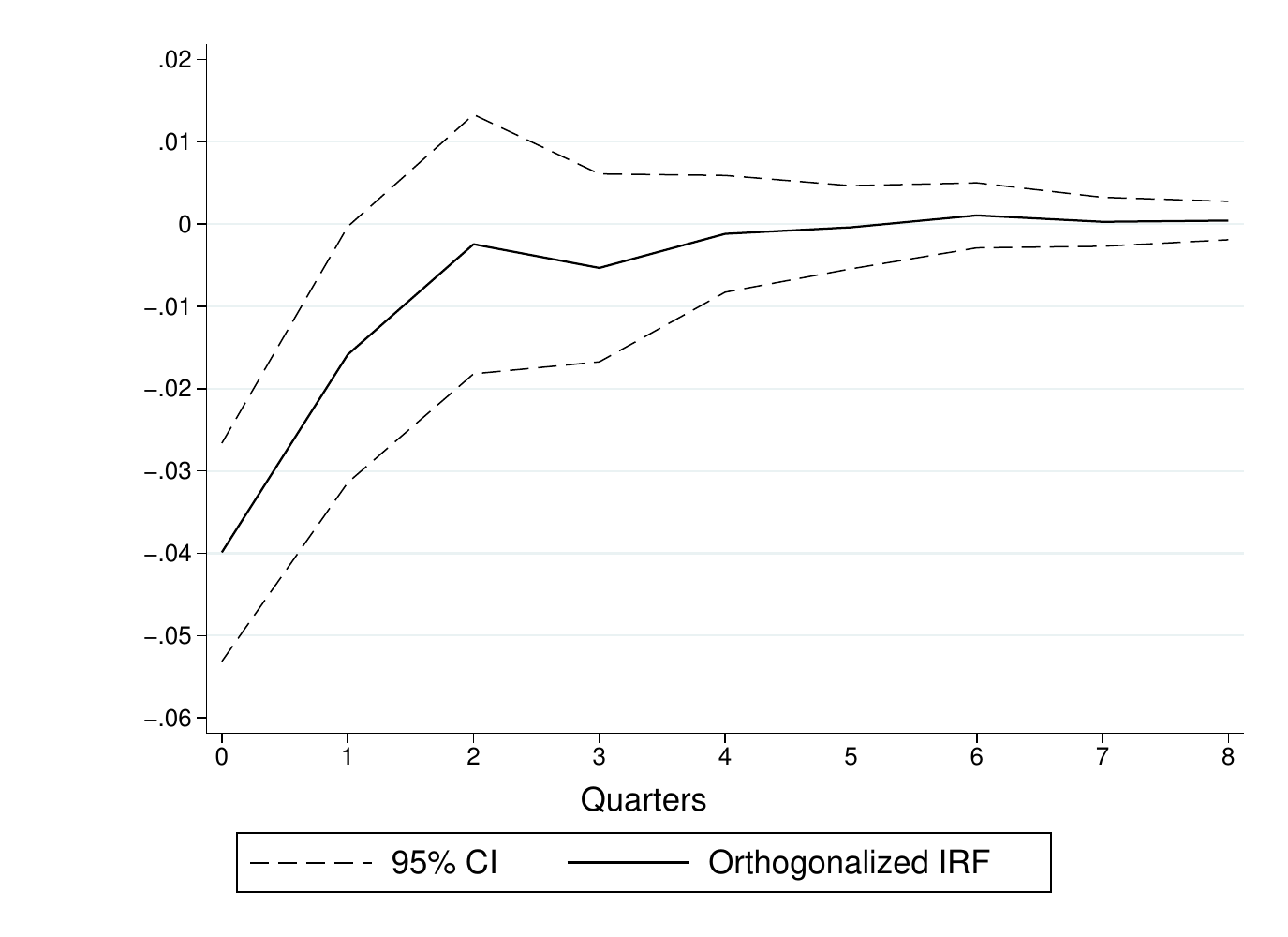}
	\subcaption{Response of exchange rate to oil shocks}
	\includegraphics[width=65mm]{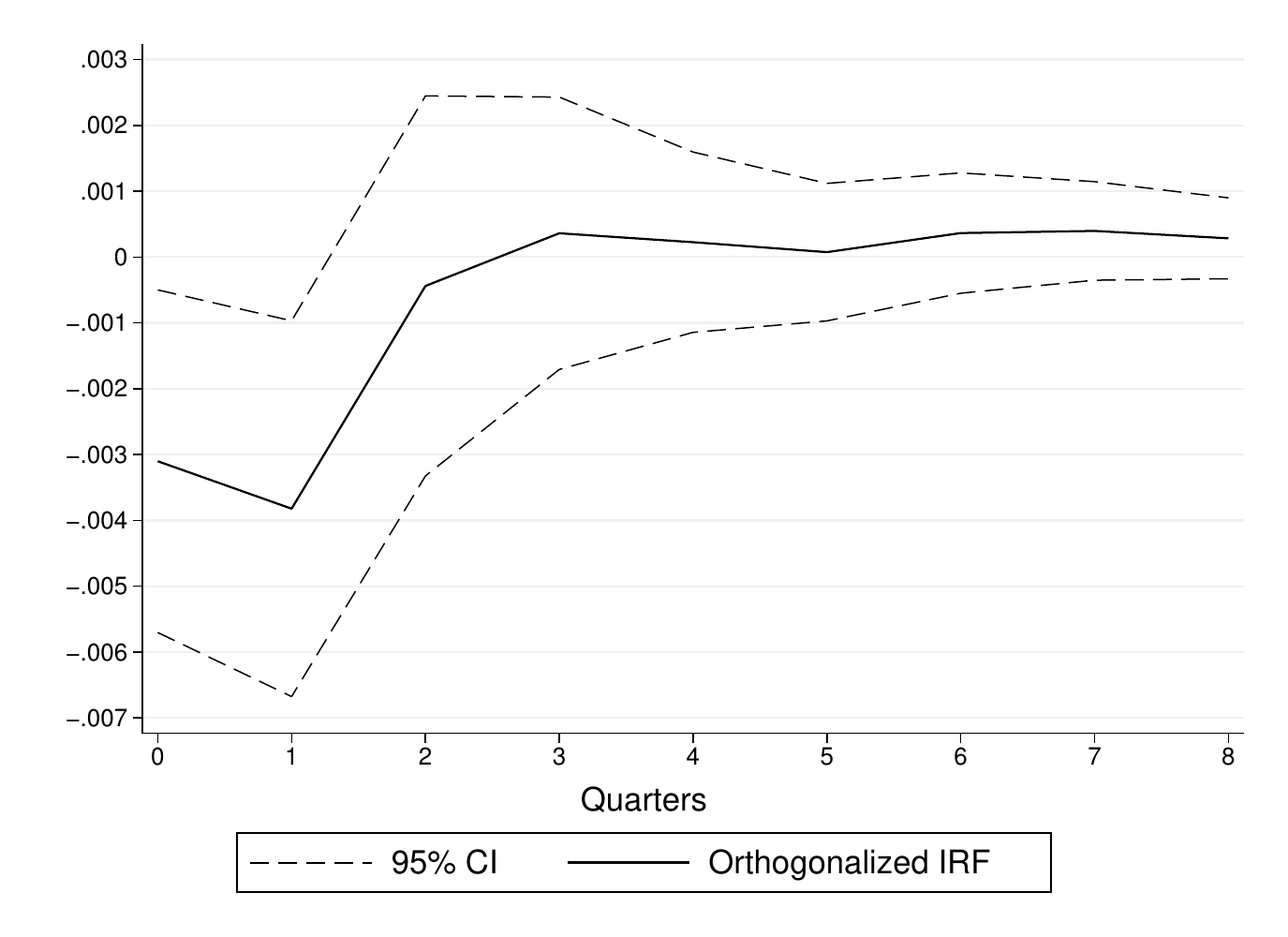}
	\subcaption{Response of inflation rate to oil shocks}
	\includegraphics[width=75mm]{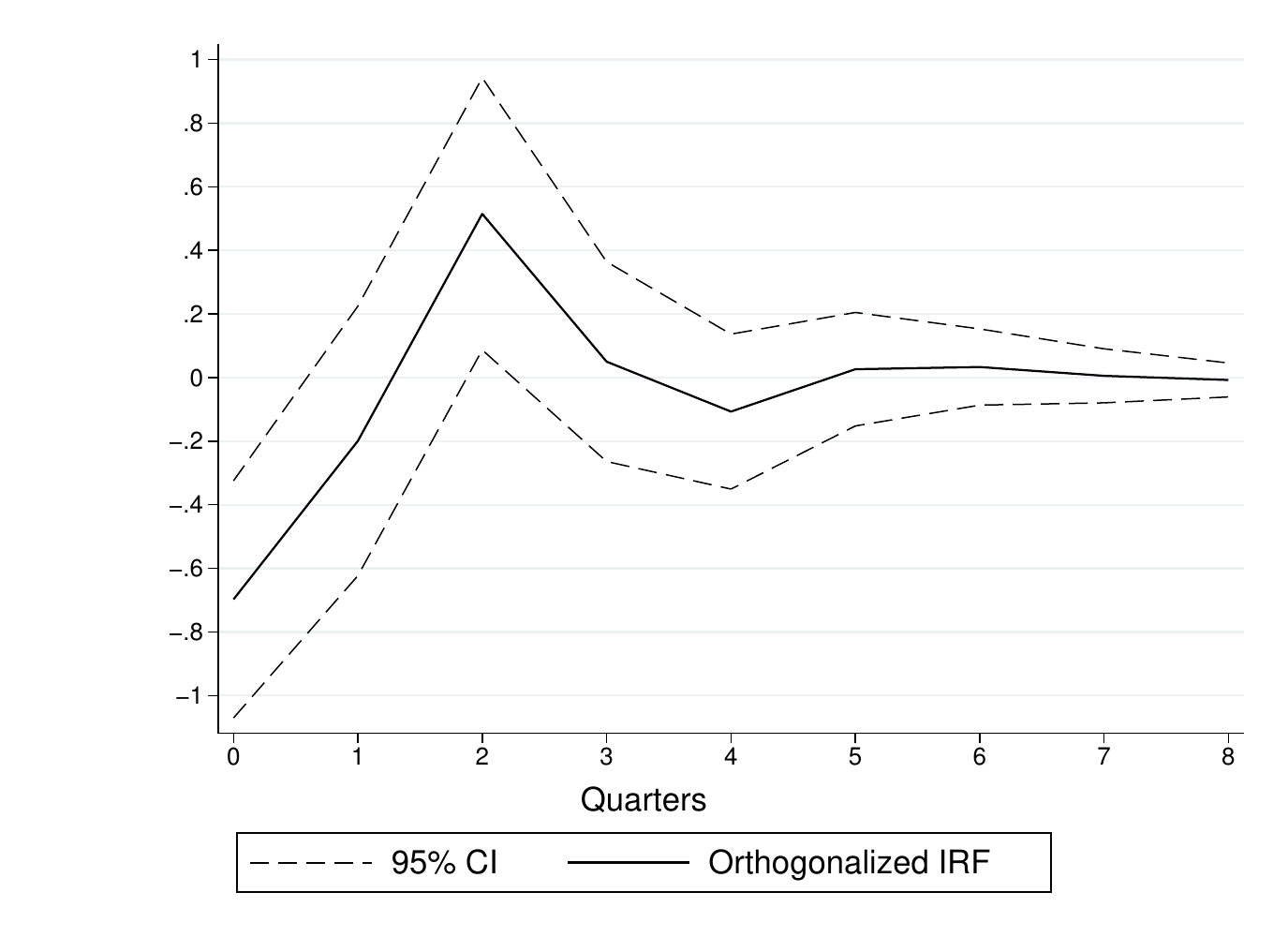}
	\subcaption{Response of interest rate to oil shocks}
	\begin{tabular*}{0.45\textwidth}{@{\hskip\tabcolsep\extracolsep\fill}ccccc}
	\multicolumn{4}{p{0.42\textwidth}}{\scriptsize \emph{Note:} Time series covers period of Q1 2000 – Q3 2021 in first differences of logarithmic form. Impulses for 8 quarters are presented.}\\ 
	\end{tabular*}
	\end{center}
\end{figure}

Figure-4(a) depicts the response of the industrial production index to the oil price shock. According to the IRF outcome, Russia's production index is positively affected by oil prices. Thus, if there is an increase in oil pricing, the production output of Russian industry sectors also increases. The results show that a 1\% increase in oil prices leads to a 0.6\% increase in the industrial production index (IPI). However, this effect is significant only for one period. After that, at a 95\% significance level, the impact is insignificant, with the confidence interval reaching the level of zero. In conclusion, oil price shock has a positive effect on the Russian IPI for the first period after it occurred.

As oil prices go up, Russian oil producers' revenues surge, creating an occasion to increase production and export values or generate possibilities of constructing new drilling or refining facilities. Moreover, as the Russian federal budget's profits increase, so do the federal exchange reserves, which appreciate the domestic currency. This strengthening of the ruble makes imports cost less for the Russian manufacturers. As it has already been discussed, the most imported items are mechanical and electrical machinery, which contribute to the Russian industries. As their costs become more affordable, it facilitates the procurement of the necessary equipment, significantly raising the production capacities of the industries and, thus, increasing the production index.

The response of the exchange rate to a shock in the oil prices is negative, indicating that with an increase in oil prices of 1 \%, the Russian domestic currency would appreciate by more than 4 \% (Figure-4(b)). The prices of oil in the world market are determined in US dollars, making it a dominant currency in which almost all deals are settled. A positive oil price shock increases the foreign exchange reserves of Russia, which for its part, strengthens the domestic currency, resulting in its appreciation. This relationship is in line with our findings, yet this negative relationship is significant solely for the first period or the first quarter after the oil price shock occurs. After that, the confidence interval reaches zero, making the impact insignificant.

The response in the inflation rate to an oil price shock is also significant for one initial period, and the relationship is negative (Figure-4(c)). This means that with an augmentation of the world oil price by 1 \%, the price level in Russia would decrease by approximately 0.3 \%. As in the case of a positive oil price shock, the domestic currency appreciates as foreign exchange reserves experience an influx. Subsequently, imported products become more affordable for domestic consumers. Even though a surplus mainly determines its current account, Russia is highly dependent on imports of intermediates and, in particular, of the end-products, making it an import-dependent country. According to the data obtained from the Federal Customs Service of Russia, the most imported goods in value in 2021 were cars and machinery, mechanical machinery, electrical machinery, and parts of motor vehicles. It can be concluded that Russian industries that require mechanical equipment rely on foreign hardware and other goods imported to Russia. Imported products also represent a large portion of the consumer basket. With its costs lowering, it creates a high degree of competition for some domestic producers, as they are not as developed to withstand it and cannot produce at a lower price.

The relationship of world oil prices with the Russian interest rate is negative (Figure-4(d)). According to the analysis outcomes, with a 1 \% augmentation of the oil prices, the interest rates in Russia would decrease by 0.7 \%. This relationship could be linked to the impact of the oil price volatility on the inflation rate. As the oil prices and the inflation rate have a negative relationship, an oil price drop would increase the price level in Russia. By Taylor's rule, when inflation rates are higher than desired, the Central Bank should increase the interest rate to mitigate the increase in the price level in the economy.

As a robustness check, we use GDP growth instead of the industrial production index. The results are consistent with the baseline model outcome (see, Figure-A3). The response of the GDP growth to the oil price shock is positive; thus, a 1 \% increase in oil prices leads to a 7 \% increase in the GDP,  9 \% and 8 \% for the following second and third periods. After the third period, the impact is insignificant, with the confidence interval reaching zero. In conclusion, oil price shocks had a robust positive effect on the Russian GDP growth for the three quarters after they occurred.

The overall results of the IRF analysis indicate that the macroeconomic variables of the Russian Federation are vulnerable and responsive to oil price fluctuations. These relationships can also be explained by the linkages of the macroeconomic variables among themselves, as was initially assumed in our analysis. Thus, the impact of the oil price shocks on the Russian interest rate might also be explained by the effects of the shock on the inflation rates. The reaction of the price level in the economy might also be linked to the oil price’s influence on the Russian exchange rate. Nevertheless, the impact of a sudden change in oil prices in the market can be observed in all four macroeconomic indicators used in my analysis.

\section{Conclusion}
\label{sec:conclusion}

Being rich with natural resources, the Russian Federation heavily relies on its exports. Trade with raw materials represents a tremendous and indispensable part of the economy of this country. Even though Russia is one of the most significant oil producers in the world, it is not capable of influencing the world oil prices. On the contrary, the oil price fluctuations affect this country's macroeconomic performance, making Russia vulnerable to oil price shocks.

We investigated the recent fluctuations in the oil market from 2004 to 2021 and their effects on the Russian macroeconomic indicators. Thus, the contribution of our research paper is to investigate the impact of oil price fluctuations on the Russian macroeconomic indicators and measure the direction and magnitude of responses to the oil shocks.

As a result of the statistical analysis conducted, the variables used to represent the macroeconomic indicators of Russia were proven to have been vulnerable to the effects of the oil price fluctuations in the period from 2004 to 2021. The response in the IPI, an indicator that represents the Russian business cycles and the output of the county's total industry, has been proven to be positive. The findings indicate that with an increase in the oil prices by one percentage, there is also an augmentation of 0.6 \% in the real output of the industrial production in the Russian economy. The energy, mining, and manufacturing sectors experience an increase in their production. On the other hand, if the oil market experiences a price shock, the industries' total output decreases. This leads to the conclusion that the Russian business cycles are vulnerable to the oil price evolution on the market.

The relationship between the oil price fluctuations and the inflation, exchange and interest rates has been proven to be negative. Thus, if there is a one percentage oil price drop in the market, the Russian domestic currency appreciates by more than four percent, and the price level increases by 0.3 \%, making the Central Bank of Russia increase the interest rates, which would eventually rise for 0.7 \% of its value. A vulnerability of the Russian domestic currency to the energy market volatility creates a threat to the sectors of the economy, which heavily rely on imported goods and services. With the prices of foreign goods increasing, so does the price level in Russia, which is dependent on imports to a significant extent.

Proceeding from all of the above, it is evident that the Russian Federation should mitigate its economic dependence on the exports of its natural resources. Diversifying the federal budget revenues should be considered to decrease the large share of the oil and gas revenues section in its incomes. Providing support to other sectors and industries in the economy would decrease Russia's reliance on producing its raw materials and reduce the Dutch disease symptoms. The large amounts of oil and gas revenues should not only serve as a safety cushion for times of economic recessions but also be a valuable source of investments in other non-petroleum assets.

\bibliographystyle{elsarticle-harv} 
\bibliography{Literature}

\newpage

\section*{Appendix}
\addcontentsline{toc}{section}{Appendix}
\label{sec:appendix}

\renewcommand{\thesection}{A\arabic{section}}%
\renewcommand{\thetable}{A\arabic{table}}%
\renewcommand{\thefigure}{A\arabic{figure}}%
\renewcommand{\theequation}{A\arabic{equation}}%
\setcounter{equation}{0}%
\setcounter{table}{0}%
\setcounter{figure}{0}

\begin{figure}[ht]
	\caption{Federal budget revenues of the Russian Federation in 2021 (by sections)}
	\begin{center}
		\includegraphics[width=70mm]{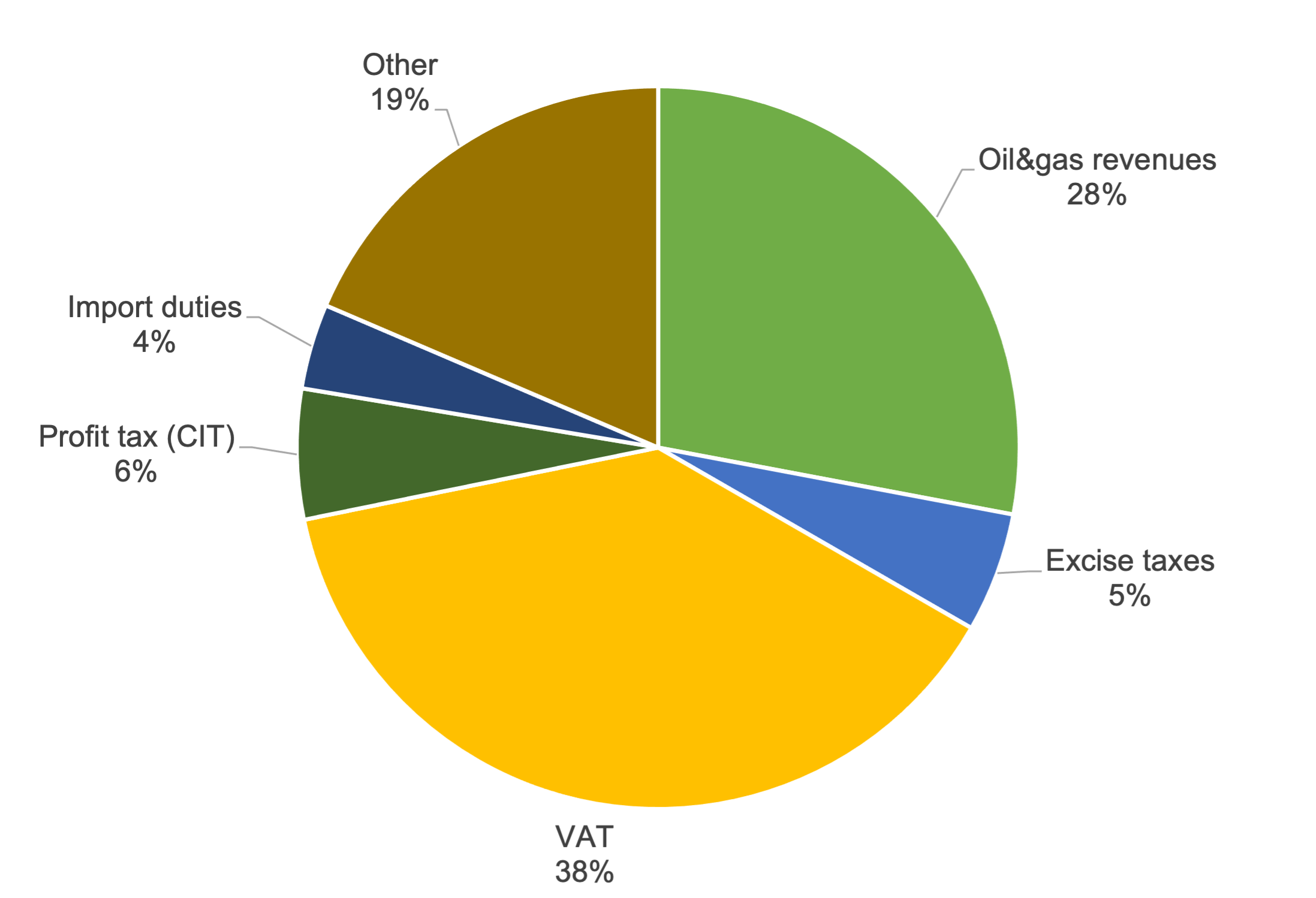}\label{fig:a1}
	\end{center}
	\vspace*{-5mm}
    	{\footnotesize Note: The authors' estimation based on statistics from the Ministry of Finance of the Russian Federation.}
\end{figure}

\begin{figure}[ht]
	\caption{Oil-related exports of the Russian Federation in 2021}
	\begin{center}
		\includegraphics[width=70mm]{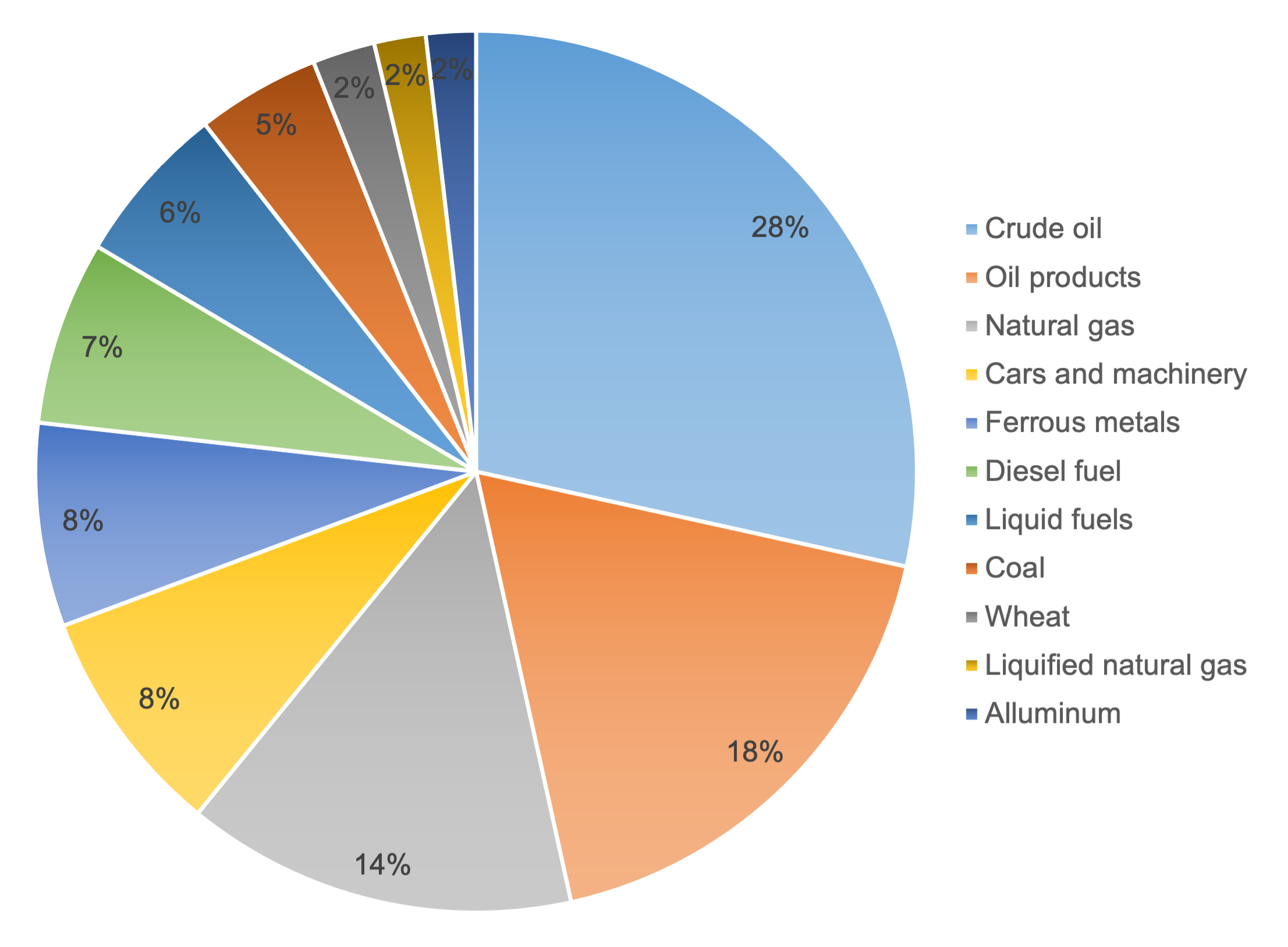}\label{fig:a2}
	\end{center}
	\vspace*{-5mm}
    	{\footnotesize Note: The authors' estimation based on statistics from the Ministry of Finance of the Russian Federation.}
\end{figure}

\begin{figure}[ht]
	\caption{Response of Russian macroeconomic indicators to oil shocks\label{fig:a3}}
		\begin{center}
	 		\includegraphics[width=60mm]{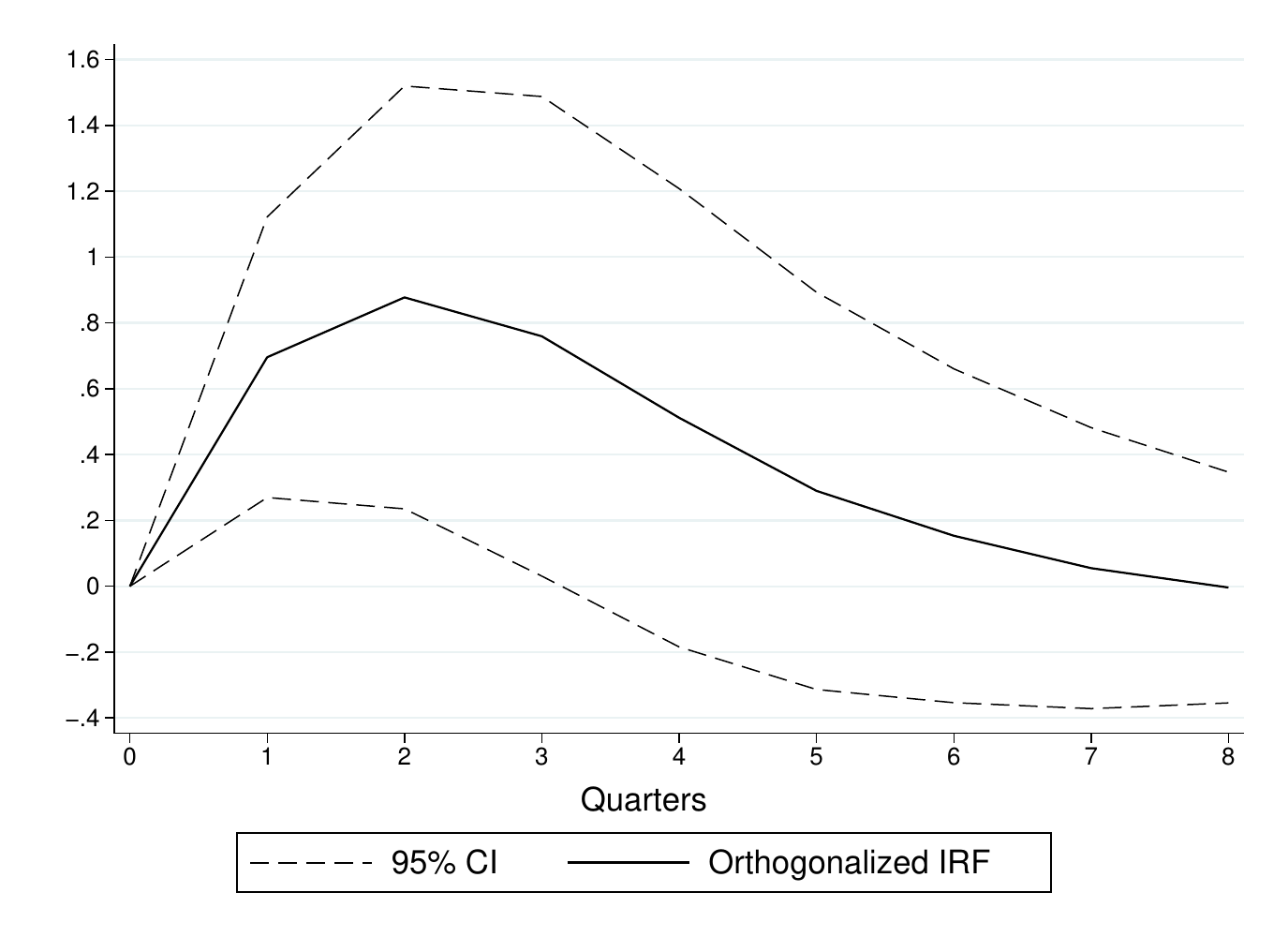}
	 		\subcaption{Response of economic growth to oil shocks}
	 		\includegraphics[width=60mm]{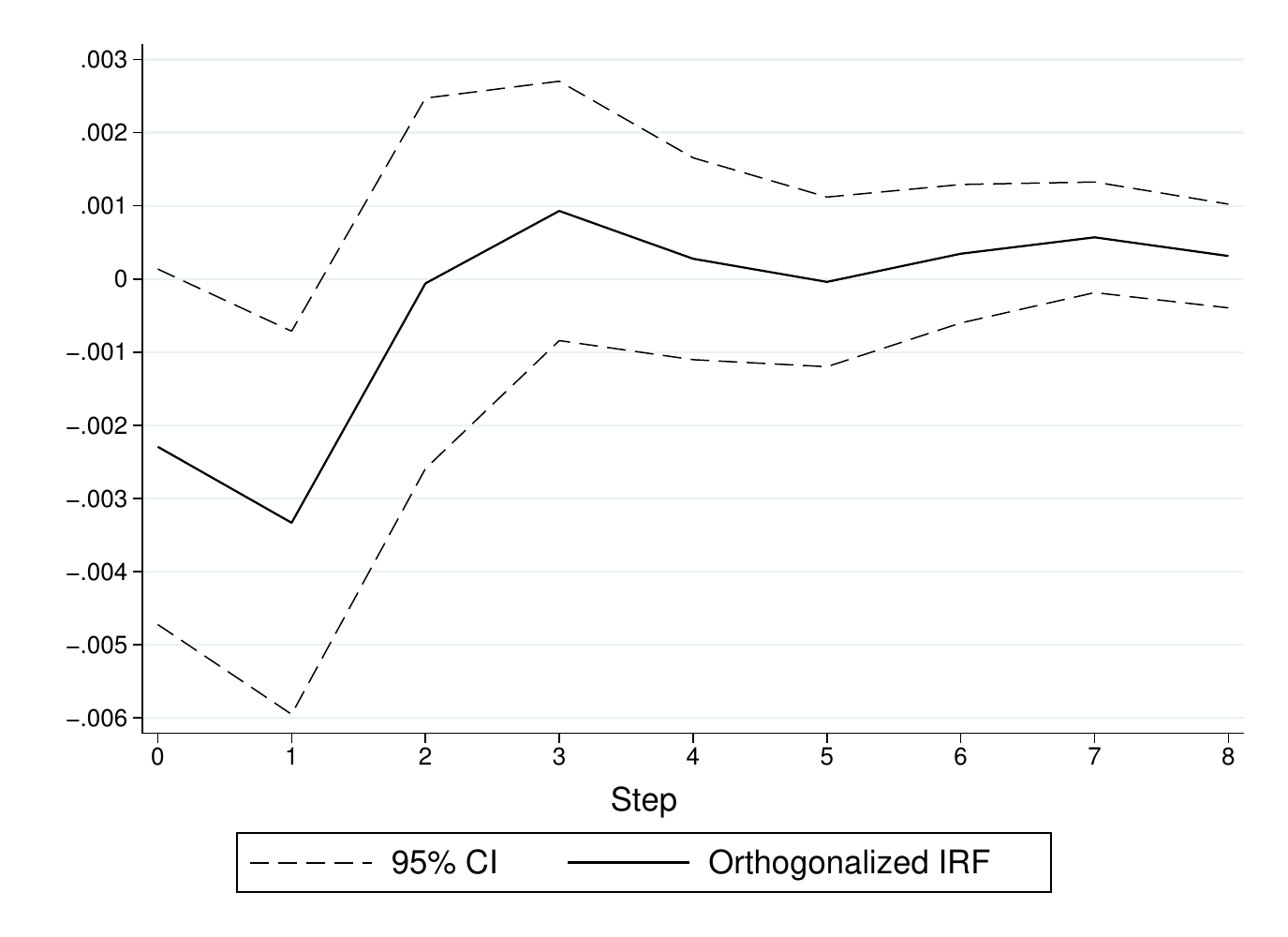}
	 		\subcaption{Response of exchange rate to oil shocks}
	 		\includegraphics[width=70mm]{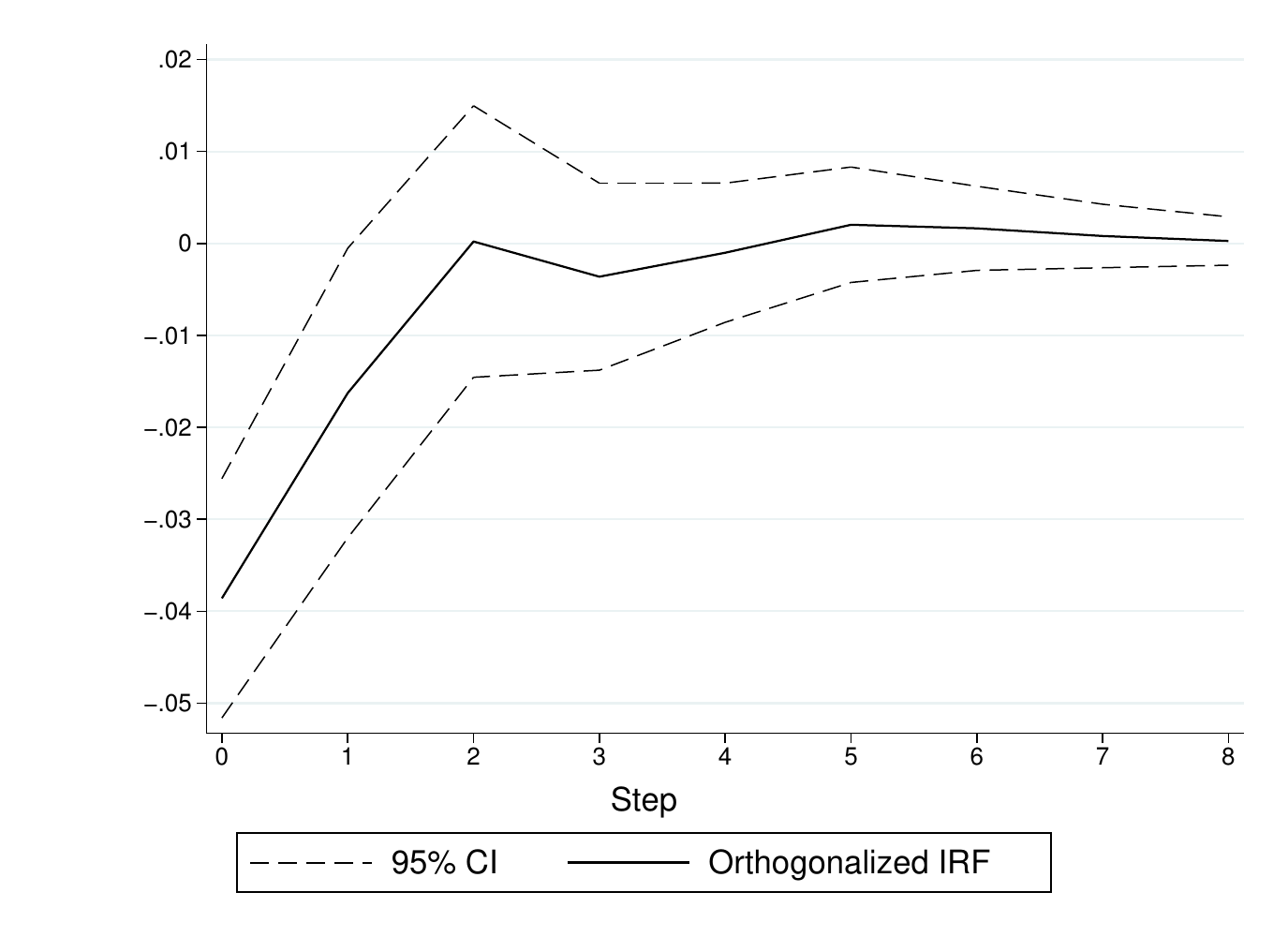}
	 		\subcaption{Response of inflation rate to oil shocks}
	 		\includegraphics[width=70mm]{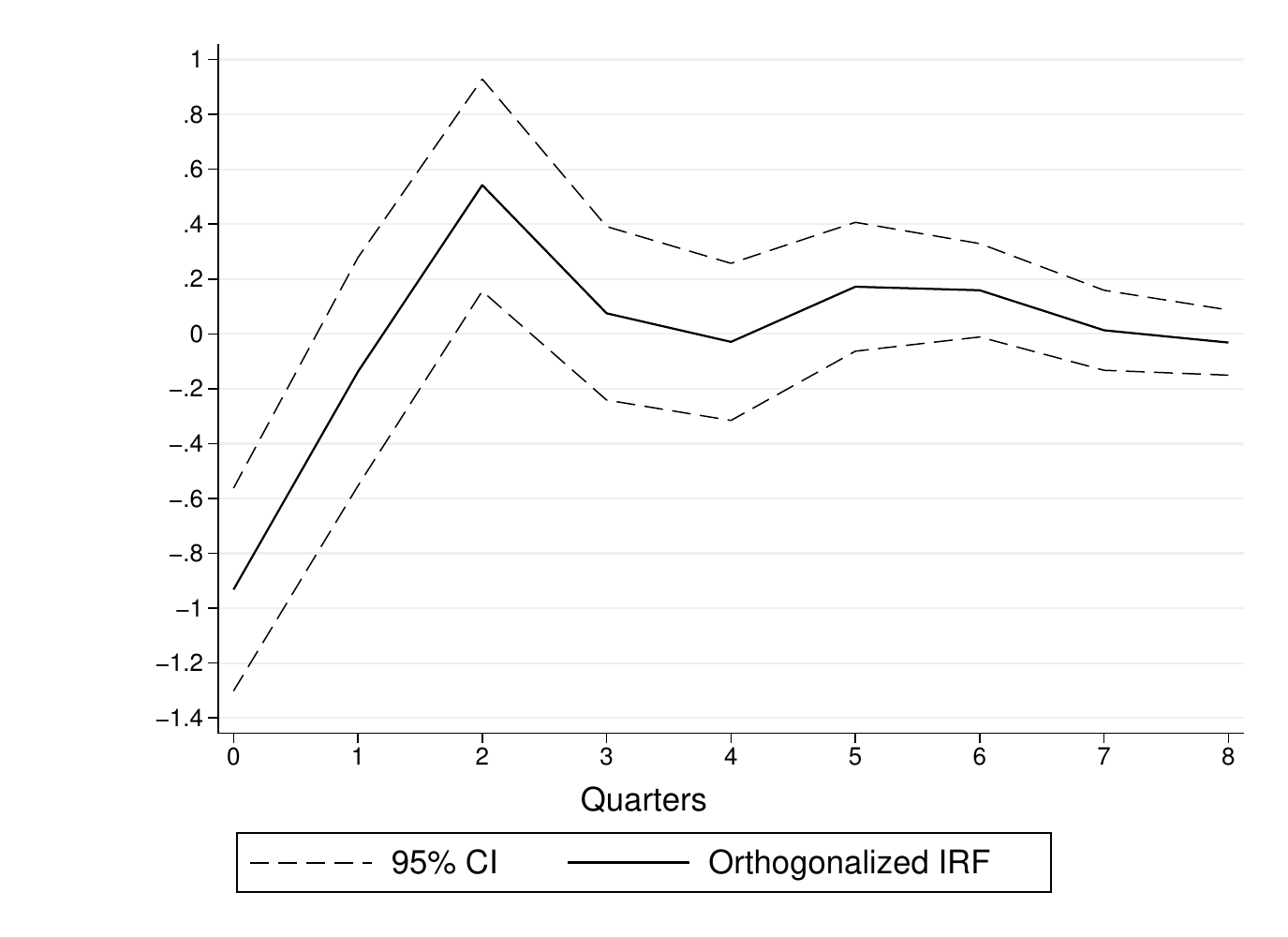}
	 		\subcaption{Response of interest rate to oil shocks}
	 	\begin{tabular*}{0.45\textwidth}{@{\hskip\tabcolsep\extracolsep\fill}ccccc}
	 			\multicolumn{4}{p{0.42\textwidth}}{\scriptsize \emph{Note:} Time series covers period of Q1 2000 – Q3 2021 in first differences of logarithmic form. Impulses for 8 quarters are presented.}\\ 
	 		\end{tabular*}
 		\end{center}
\end{figure}

\end{document}